# The role of temperature on defect diffusion and nanoscale patterning in graphene

Ondrej Dyck,[1,*] Sinchul Yeom,[2] Sarah Dillender,[3] Andrew R. Lupini,[1] Mina Yoon,[2] Stephen Jesse[1]

[1] *Center for Nanophase Materials Sciences, Oak Ridge National Laboratory, Oak Ridge, TN, USA*

[2] *Materials Science and Technology Division, Oak Ridge National Laboratory, Oak Ridge, TN USA*

[3] *Princeton University, Princeton, NJ, USA*

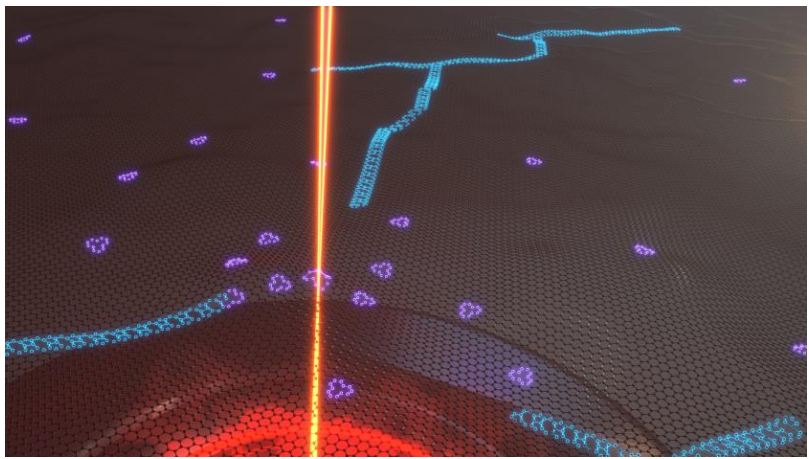

**Abstract**

Graphene is of great scientific interest due to a variety of unique properties such as ballistic transport, spin selectivity, the quantum hall effect, and other quantum properties. Nanopatterning and atomic scale modifications of graphene are expected to enable further control over its intrinsic properties, providing ways to tune the electronic properties through geometric and strain effects, introduce edge states and other local or extended topological defects, and sculpt circuit paths. The focused beam of a scanning transmission electron microscope (STEM) can be used to remove atoms, enabling milling, doping, and deposition. Utilization of a STEM as an atomic scale fabrication platform is increasing; however, a detailed understanding of beam-induced processes and the subsequent cascade of aftereffects is lacking. Here, we examine the electron beam effects on atomically clean graphene at a variety of temperatures ranging from 400 to 1000 ºC. We find that temperature plays a significant role in the milling rate and moderates competing processes of carbon adatom coalescence, graphene healing, and the diffusion (and recombination) of defects. The results of this work can be applied to a wider range of 2D materials and introduce better understanding of defect evolution in graphite and other bulk layered materials.

---

[*] Corresponding Author. E-mail: dyckoe@ornl.gov (Ondrej Dyck)

## 1 Introduction

Graphene exhibits a number of scientifically interesting and technologically relevant properties that can be further enhanced when its structure is appropriately altered. Spin and band gap engineering may be performed by doping graphene nanoribbons (GNRs) with various elements.[1] The width and edge terminations of GNRs can be used to tune the electronic properties.[2] Together, these variations open doors for their use as quantum-based devices such as spin diodes,[3–6] quantum dots and spin qubits,[7–10] and ballistic rectification and transport.[11–14] These applications require tailoring graphene with atomic-scale precision, which is beyond current nanofabrication capabilities.

The first step toward this goal is obtaining and maintaining atomically clean graphene, which has proven to be a challenge, as immediately after growth, contaminants begin to adhere to the surface.[15] Nevertheless, several studies have shown how clean graphene may be attained.[16–19] Of particular interest are strategies that produce large areas of clean graphene such as rapid thermal shock—introduced through laser or joule heating that may be compatible with wafer scale device fabrication.

Once atomically clean graphene can be reliably obtained and sustained, the next step in the fabrication workflow is to add and remove atoms from the lattice in a controlled way to imbue the substrate with the desired functionality. One strategy that may offer a path forward is utilizing a scanning transmission electron microscope (STEM) as a fabrication platform.[20] Demonstrations along these lines have included graphene patterning and deposition at the single-digit nanometer scale.[17,21–29] Efforts to scale deposition and patterning to the scale of single atoms requires careful consideration of atomic processes. For example, e-beam induced deposition (EBID) is typically conceptualized as a dissociated precursor gas attaching to a substrate without atomic level specificity.[1] Attempting to scale this concept to a single atom one runs into trouble as there is no precursor to dissociate, only single atoms, and thus there is no mechanism for directed deposition. Moreover, a single atom deposited on an atomically pristine surface is merely an adatom. The activation energy for the diffusion of adatoms on a pristine surface is low enough that for many practical purposes the precise positioning of adatoms would be fruitless. Both of these challenges can be addressed, however, if the e-beam is not used to chemically alter the precursor but instead used to chemically alter the substrate. At the single atom level, this requires that the substrate harbor a beam-induced vacancy/defect/impurity to facilitate stronger chemical bonding. This chemical bond location cannot exist uniformly across the sample surface or the deposition will no longer be localized. Thus, to deposit a strong, chemically bonded single atom at a specific location requires both the generation of a localized chemical bonding site and the delivery of the single atom to that location. Investigations along these lines began with Si atoms introduced into e-beam generated defect sites in graphene.[30,31] The same procedure was shown to facilitate the attachment of many other atomic species to e-beam generated graphene defect sites, illustrating its general applicability.[32–34] This approach requires the ejection of one or several atoms from the graphene lattice using an e-beam with an energy above the atomic displacement (knock-on) threshold.

Investigations into the atomic displacement process have found a role for thermally assisted damage.[35–37] The knock-on process, i.e. the direct transfer of energy and momentum from the incident electron to the atomic nucleus, is of singular importance because it is the primary mechanism used to drive the generation of atomic vacancies that enable further alterations.

Increasing the sample temperature increases the likelihood of ejecting an atom for a given e-beam energy resulting in a softening of the sharp displacement threshold onset of the McKinley-Feshbach description, which assumes a static lattice.[37,38] Given this understanding, one would expect the overall effect of increased sample temperature to result in more rapid degradation of graphene under the STEM e-beam. Such an effect could be used, for example, to tune the likelihood of removing an atom from the lattice when the accelerating voltage is close to the knock-on threshold. Here, we test this hypothesis and find that spontaneous healing[39] and vacancy diffusion[40] have pronounced, dominant, and counterintuitive effects that upend the original hypothesis. Understanding and controlling these processes represents a critical step in the development of fabrication workflows that can be refined toward atomic precision.

## 2   Material and methods

### 2.1   Experimental Strategy

The ejection of C atoms from a graphene lattice with an e-beam is a statistical process. To properly capture and study such processes requires a repeatable experiment that can be performed iteratively, with minimal user input, to obtain sufficient statistical sampling. To accomplish this goal, we used a custom beam position control and signal processing platform that interfaces with the microscope to allow arbitrary scan paths and automated experimental control via feedback loops.[41–43] This platform enables the sequential creation of small holes in the lattice of single-layer suspended graphene under consistent experimental parameters, from which statistical information regarding beam-induced changes are gathered and analyzed. Figure 1 shows a diagrammatic representation of the experimental approach. Standard, raster-scanned images are acquired to obtain overview images of the state of the sample, in what we refer to as "imaging mode" (Figure 1(a)). Within the field of view (FOV), an array pattern is chosen that determines the locations of holes to be milled, as shown in Figure 1(b). To mill a hole, the scanning pattern is changed to the "milling mode," as shown in Figure 1(c). In this mode, the e-beam is scanned in a small spiral (approximately 3 – 10 lattice spacings in diameter). This scan trajectory was chosen for several reasons. The smooth, single-frequency waveforms comprising the x(t) and y(t) drive signals sent to the scan coils have little to no sharp transitions and allow for relatively fast, distortion-free imaging as compared to traditional raster scans. Also, for a given scan, the e-beam spends a disproportionate amount of time at the center of the scan, resulting in a higher fluence and accumulated dose, thereby strongly increasing the likelihood of local vacancy generation while still allowing tracking of changes in the atomic structure in the immediate neighborhood. The mean intensity of the medium angle annular dark field (MAADF) detector signal over a spiral scan is used as the primary input to the feedback system and a user-defined threshold is set to determine when a hole is formed. Due to changes in e-beam current over time, different beam current settings, or different detector gain settings, the mean intensity of the graphene can vary significantly, requiring adjustment of the threshold. Figure 1(d) illustrates the milling process over time. The data points represent the mean spiral scan image intensity as a function of time. As defects accumulate under the scan location, the mean intensity drops. When the mean drops below the user-defined threshold, the beam is automatically advanced to the next milling location and the process is repeated. Example images are shown across the top, demonstrating the formation of a hole and the e-beam advancing to the next location.

It is useful to carefully consider the atomic level details of this process. We mentioned that the 100 kV e-beam creates vacancies through a knock-on process. For purposes here, we do not consider a single vacancy as a "hole" and reserve this term to mean a larger region of missing lattice. If holes could be formed by the targeted sequential ejection of carbon atoms from a static lattice, this distinction would be unnecessary; however, in the real world, the graphene lattice is never static under e-beam irradiation and the sequential ejection of carbon atoms occurs simultaneously with lattice restructuring. Many carbon atoms can be removed from a graphene lattice resulting in a reconstructed defect cluster without a clearly defined hole appearing. A hole can only be formed if the local *density* of vacancies reaches some critical value. For example, one can imagine the case of adding single vacancies indefinitely to a defect cluster in such a manner as to form an ever-larger defected region, without ever forming a clear hole, i.e., without the defect density reaching the critical value.

Thus, the milling of a hole is separated into the following stages: 1) vacancies are created by the ejection of carbon atoms from the graphene lattice, 2) the lattice restructures, 3) the defective region expands as additional vacancies are generated, 4) the ejection of an additional carbon atom from an already defected region increases the vacancy density to such an extent that it is locally more energetically favorable to restructure into a hole, 5) the graphene lattice undergoes a substantial restructuring, vacancies coalesce, and a hole is created with the edge harboring many dangling bonds.

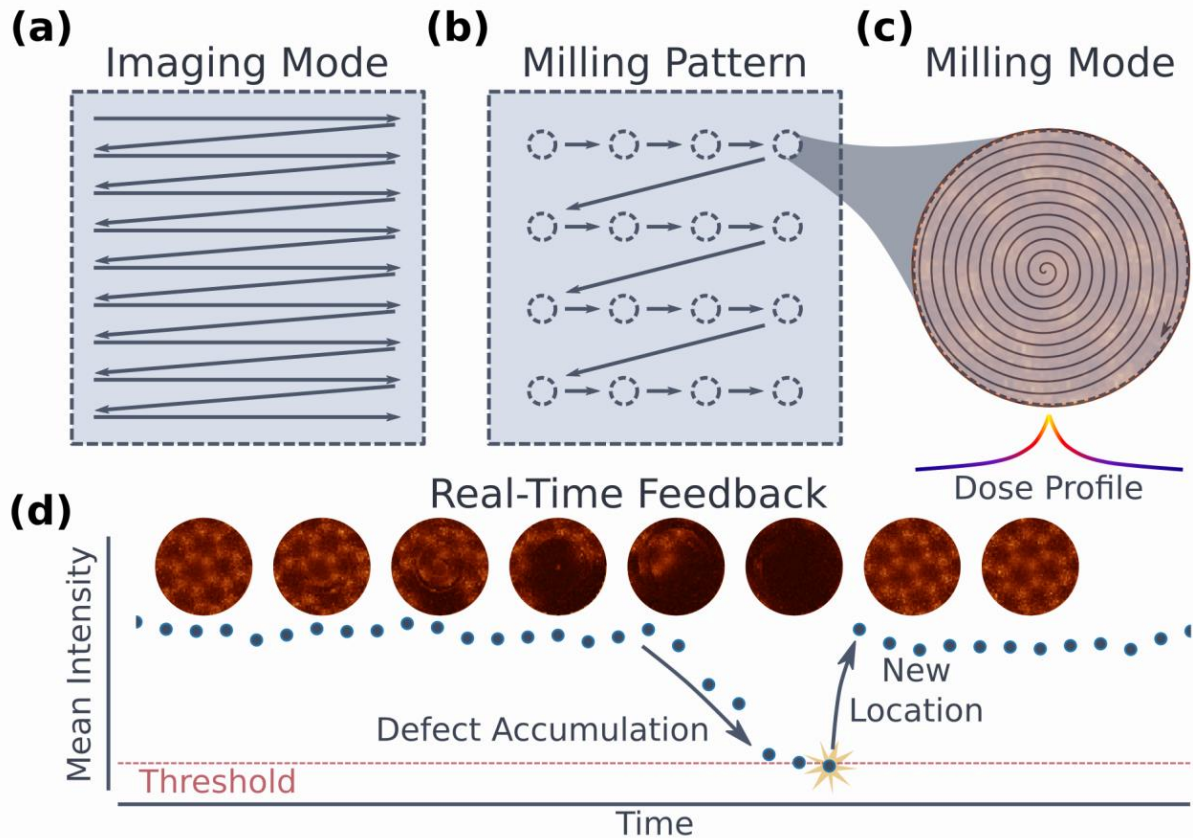

**Figure 1 Illustration of the experimental strategy.** (a) In "imaging mode" a standard raster scan is performed to obtain an overview of the sample. (b) illustration of the target array pattern. (c) In "milling mode" a small spiral scan is performed with the dose concentrated at the center of the spiral. (d) Illustration of the spiral scan mean intensity across time. As defects accumulate in the scan area the mean intensity drops. When it drops below the threshold the e-beam is advanced to the next milling location. Example spiral scan images acquired during a milling cycle are shown across the top.

## 2.2 Experimental Preparation

Graphene was grown on Cu foil using chemical vapor deposition (CVD). The graphene was capped by spin-coating with poly(methyl methacrylate) (PMMA). To transfer the graphene to a STEM compatible substrate, the Cu/graphene/PMMA stack was then floated on a solution of ammonium persulfate and deionized (DI) water to dissolve the Cu. Once the Cu foil was dissolved, the graphene/PMMA stack was rinsed in DI water and caught on a Protochips Fusion heater chip. The sample was then dried on a hot plate at 150 ºC for 15 minutes. Once cooled, the heater chip was immersed in acetone to dissolve the PMMA. Finally, the chip was dipped in isopropyl alcohol to remove the acetone and allowed to air dry. To remove residue acquired during the transfer process, the heater chip was ramped to 1200 ºC in vacuum at a rate of 1000 ºC/ms. Finally, the sample was baked in vacuum at 160 ºC for 10 hours prior to loading in the microscope.

A Nion UltraSTEM US200 was used for these experiments and was operated at an accelerating voltage of 100 kV. Images were acquired using the medium angle annular dark field (MAADF) detector. The nominal convergence angle was 30 mrad. The beam current was in the range 50-65 pA except for the images shown in Figure 2(b) where the beam current was decreased to ~20 pA to improve resolution.

After insertion in the microscope the sample was ramped to 900 ºC at a rate of 1000 ºC/s to remove any residual contamination. Images acquired of the initial sample at room temperature and after the rapid heat treatment can be found in the Supplemental Information (SI). Significant e-beam induced hydrocarbon deposition was observed after heating the sample. In a previous publication we described a strategy for mitigating the ingress of such hydrocarbons through the *in situ* deposition of physical barriers.[44] Here, we made use of this strategy to maintain clean areas of graphene for the experiments. The sample examined here is the same sample described previously;

images as well as a detailed description of the deposited barriers and their effectiveness at reducing deposition are also presented.[61]

## 3 Results and Discussion

Using the strategy described above, many arrays of holes were milled in the graphene at a variety of different temperatures ranging from 400 to 1000 ºC. The electron dose required to mill each hole was recorded and the mean dose is plotted as a function of the graphene temperature in Figure 2(a). The error bars represent the standard deviation. Representative examples of overview images of several milled arrays are shown overlaid. While these arrays appear incomplete, each hole was successfully milled except for the 1000 ºC experiment and 400 ºC experiment, as will be discussed in more detail below. Due to the small size of the holes, C adatom migration and reintegration into the lattice at hole edges leads to "healing"[39] or refilling of the hole with structurally defected graphene and, particularly at the lower temperatures, many of the milled holes are no longer observed after completion of the array. See SI for a video of an array being milled while holes are concurrently healing.

What is immediately notable is the significant increase in dose required to mill a hole as the temperature increases. For the 1000 ºC experiment, we limited the number of spiral scan passes for each milling attempt to 200. Of the 25 milling attempts, only 11 holes were successfully milled for this number of spiral scan passes. At all the other temperatures every hole was able to be milled using less than 200 attempts. The dose value shown for experiments at 1000 ºC is calculated assuming the unsuccessfully milled holes were successful on the 200$^{th}$ attempt, at a dose of ~180x10$^9$ e/nm$^2$. Since this assumption is not correct, the true mill time (dose) at 1000 ºC lies somewhere above the point indicated in Figure 2(a).

This observed phenomenon, increasing resistance to damage at elevated temperatures, is remarkable. As another example, we acquired a twenty-frame stack of images at 900 ºC and aligned and summed the image stack to produce the artificially colored image shown in Figure 2(b). This image shows the graphene lattice under a 100 kV e-beam acquired with a dose of $10^{10}$ e/nm$^2$. No significant changes in the graphene were observed. In contrast, the gray scale image shows the final frame from a 45-frame image sequence acquired at room temperature. Point defects were observed after an exposure of ~$10^8$ e/nm$^2$ shown in the SI. At a dose of $3\times10^9$ e/nm$^2$ at room temperature, there is almost no graphene left in the exposed area.

Thus, two distinct behaviors are observed: At the lower end of the temperature range (400 ºC, which is still hot compared to room temperature), the holes are easy to mill but tend to be filled in by mobile contamination, whereas at higher temperatures, the graphene seems to locally resist damage, but once milled the holes are less likely to be filled in.

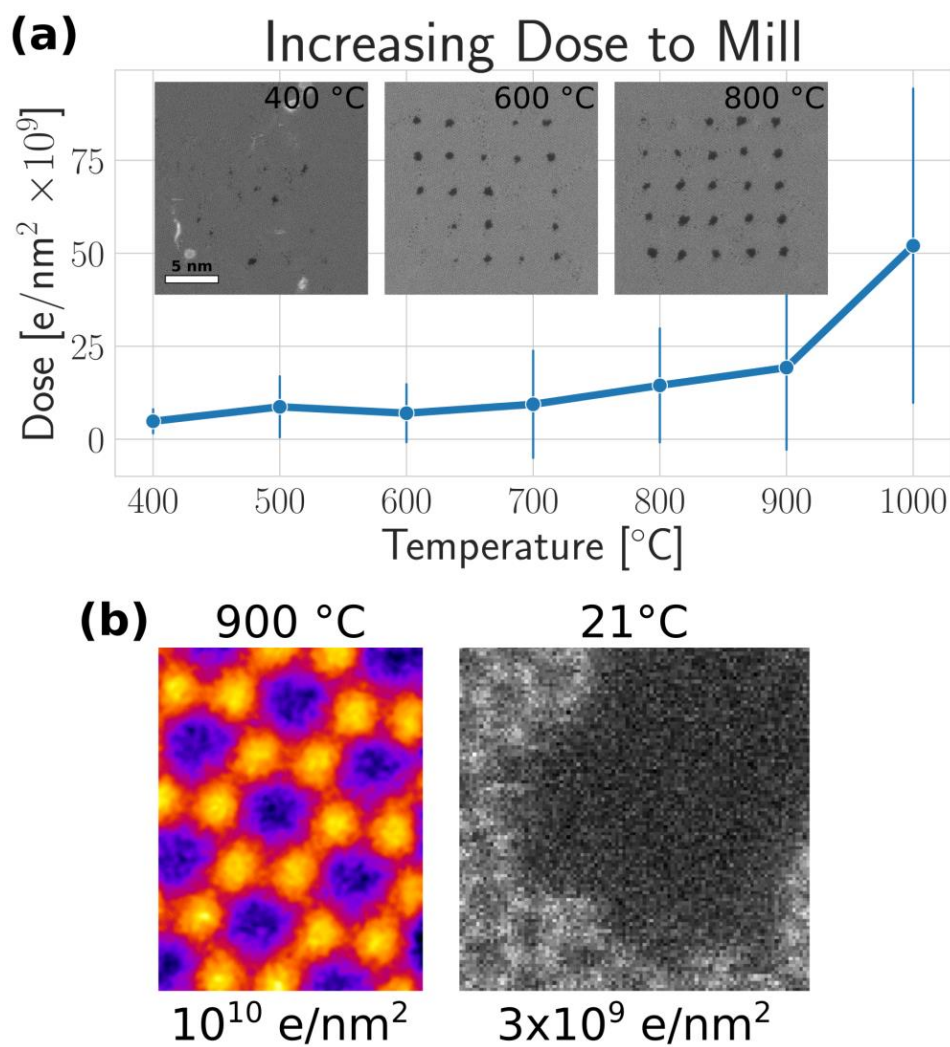

**Figure 2 Summary statistics of milling rate with temperature and example imaging above the knock-on threshold for graphene.** (a) Plot of mean electron dose needed to mill a hole in graphene as a function of temperature. (b) example image acquired at 900 °C illustrating the ability to obtain damage-free images above the knock-on energy for graphene at elevated temperatures. It has been artificially colored using the "Fire" look up table in Fiji.[45] The gray scale image shows the state of a graphene sample at room temperature after receiving an order of magnitude less dose. The accumulated dose is listed at the bottom. The field of view is 1.6 nm.

This apparent resistance to damage at elevated temperatures is at odds with current understanding of the role of phonon excitation in *reducing* the displacement threshold energy.[36] Given that we find no faults in the logic that phonon excitation should reduce the displacement threshold energy, we seek an explanatory framework that allows for an increased damage rate with increasing temperature while simultaneously providing an explanation for the observed decrease in local damage. These constraints appear to be incompatible with each other at first glance. However, we propose that two mitigating processes are of primary importance for the effective milling rate of graphene: 1) the rate of accumulation of C adatoms and molecules that attach to defect sites and heal them and 2) the rate of defect migration and accumulation elsewhere within the graphene lattice. The former process appears to be responsible for the more pronounced healing rate observed at lower temperatures (400 and 500 ºC). In particular, we note that experiments are usually difficult to perform at around 300 ºC because of mobile surface contamination. At higher temperatures the rate of hole healing was suppressed (see the example arrays shown in Figure 2(a)). This observation is consistent with the understanding that only a small area of the sample support is heated, with most of the surrounding area maintained at room temperature, acting as a sink for excess contamination. The latter process (defect migration) appears to be responsible for the increasing robustness of the graphene at higher temperatures even with an apparent lack of material with which to heal, which will be examined more closely in the rest of the manuscript.

It is worth describing explicitly the implications of this second proposed process, particularly with regard to the two images shown in Figure 2(b). Given that the displacement threshold energy decreases with increasing temperature, the graphene imaged at room temperature must have *less* accumulated damage than the graphene imaged at 900 °C (neglecting the effects of healing), which is not evident in the images. However, the images can only capture evidence of damage within the highly localized regions being sampled. If the graphene imaged at 900 °C has accumulated more total damage than the graphene imaged at room temperature, as predicted by theory, the vacancies (i.e. the fundamental unit of local damage) must diffuse away from the imaged region. Moreover, these vacancies must diffuse away at such a rapid rate that no obvious evidence of their existence is manifest in the 900 °C image. This argument implies that this image is not an image of a handful of stationary atoms in the graphene lattice, rather it captures the time averaged likelihood of occupying various spatial positions sampled by the ensemble of atoms making up the lattice. The dynamics of these processes must occur on a much shorter time scale than the imaging process. While this is a possible explanation for the observations and consistent with the current understanding of displacement threshold energies, further evidence is needed to demonstrate that this hypothesis is correct. It is the aim of the remainder of the manuscript to furnish such evidence.

A second experiment was designed to shed additional light on these processes. A single milling location was chosen at the center of the field of view (FOV) and an overview image was captured every five milling attempts until a hole was successfully created. With this strategy, the surrounding graphene could be observed periodically during the milling process. Figure 3 shows a summary of this experiment. The initial configuration is shown in Figure 3(a) and the milling location is indicated by the dotted circle at the center of the FOV. In Figure 3(b), we observe a few defects forming a chain at the edge of the image, along the zig-zag (ZZ) direction. Figure 3(c) shows the chain branching toward the milling location, again following the ZZ-direction. In Figure 3(d), a second branch has grown from the first, again toward the milling location and in the ZZ-direction. Finally, this second branch grows into the milling location and in Figure 3(e), a hole appears. A higher resolution image of one of these defect chains is shown in Figure 3(f) with an accompanying atomic model overlay. Atomic positions were extracted from the image using AI Crystallographer.[46–48] Similarly structured defects in graphene have been observed previously.[49–51] Of particular interest here is the formation of the 5-8-5 reconstructed divacancy[52–54] and growth into extended defect chains through the addition of single vacancies guided by strain fields.[55] The proposed defect chain formation process is depicted in Figure 3(g). The ejection of C atoms from the lattice occurs by the e-beam irradiation. At elevated temperatures, these vacancies rapidly diffuse through the lattice, as discussed later. When two single vacancies coalesce into a divacancy they reconstruct to the lower energy 5-8-5 structure, which induces a local strain field that acts to attract additional vacancies into the compressed regions around the 5-member rings and away from the strained regions.[55] This acts to guide diffusing vacancies to the ends of the chain and promote chain growth. Though chain branching is less favorable compared to chain extension,[55] its likelihood of occurrence does increase as the chain length increases for the simple reasons that the number of possible branching nucleation sites increases and the vacancy-attracting chain ends are farther apart. In Figure 3(c) and 3(d), branching occurred when the ends of the chain no longer pointed toward the source of the vacancies, i.e., the milling location. It should be noted that, although our focus here is on the generation of single vacancies and their coalescence into 5-8-5 reconstructed divacancies and longer defect chains, these are not the only types of defects the e-beam can generate.[56] We did not attempt to identify and categorize all forms of defect structures observed or compare the statistical likelihood of occurrence.

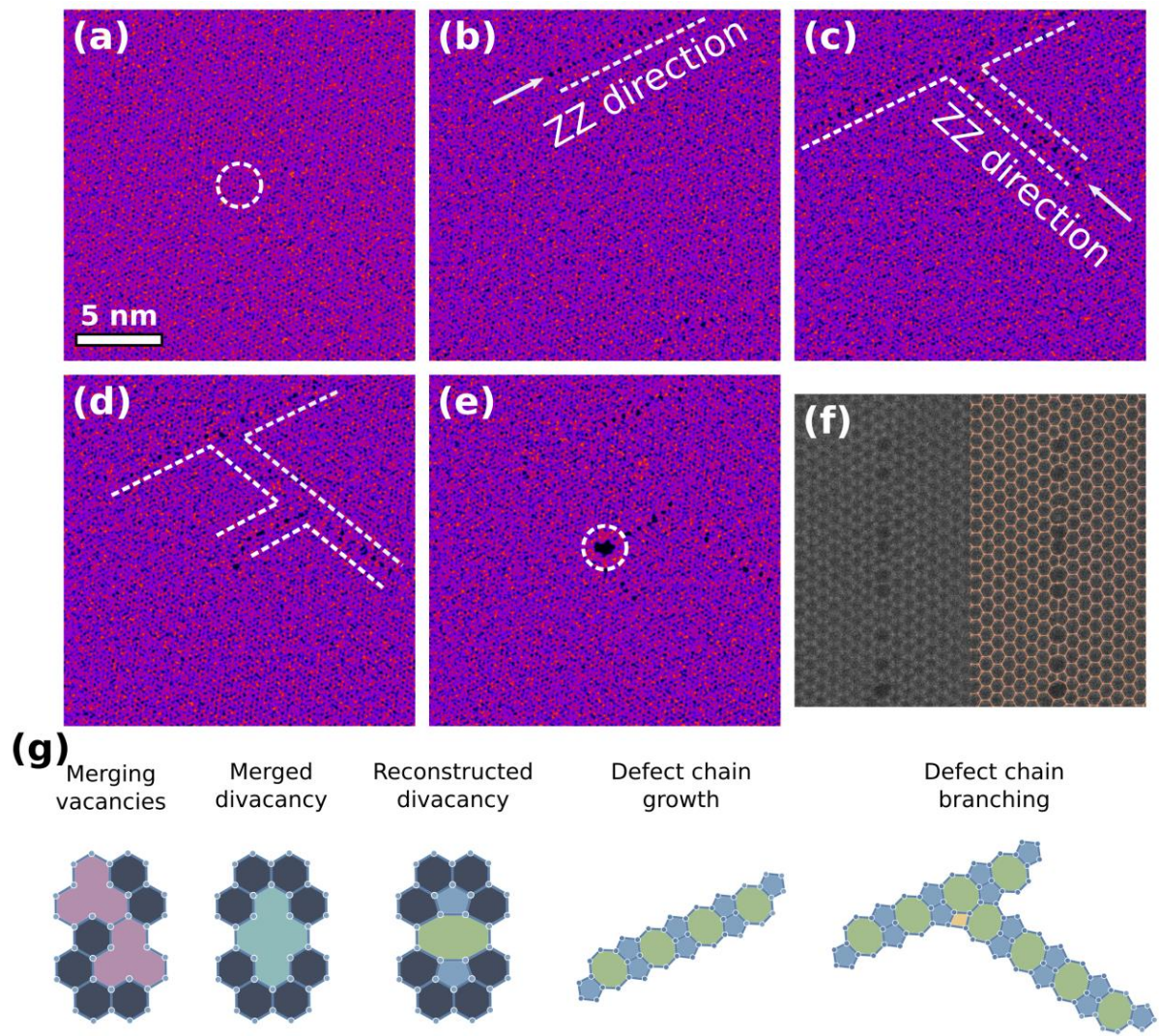

**Figure 3 Defect diffusion and defect chain growth at 900 ºC during milling.** (a)-(e) Series of images of the graphene surrounding a milling location. (a) The initial configuration with the mill location indicated at the center by the dashed circle. (b) A defect chain beginning to form along the ZZ-direction. (c) and (d) Defect chain branching. (e) Hole formation at the milling location. (f) Higher resolution image of a defect chain (left) and an atomic model overlay (right). (g) Illustration of the defect chain growth and branching process.

To put this description on a more quantitative theoretical footing, we performed classical molecular dynamics (MD) simulations based on Adaptive Intermolecular Reactive Empirical Bond Order – Morse (AIREBO-M)[57] empirical potential implemented in Large-scale Atomic/Molecular Massively Parallel (LAMMPS),[58] as summarized in Figure 4. A 92.3 x 67.6 Å$^2$ graphene sheet (2464 atoms) in a periodic boundary cell was created, the temperature was kept at 2500 K using canonical ensemble (NVT), and a carbon atom was removed nearest to the center every 10 ns. The atom trajectories were saved every 0.5 ns to reveal the structural dynamics and reordering of single vacancies into vacancy clusters.

Figure 4(a) shows the diffusion dynamics of a single vacancy. The vacancy diffuses by repeatedly breaking and forming bonds with the nearest carbon atoms. Figure 4(b) illustrates the atomistic details of two reconstructed vacancies that merge to form a reconstructed divacancy consisting of two five-member rings on either side of an eight-member ring, the topological 5-8-5 defect. This structure forms the basic building block for the observed defect chains.

Figure 4(c) shows the change in total energy during the MD simulation. The blue line tracks the energy of the system with time. Every 10 ns, the ejection of a C atom and creation of a new vacancy manifests as a discrete jump in energy. Vacancy diffusion through the pristine areas of graphene results in no net change in energy. However, restructuring of two single vacancies into a reconstructed divacancy can be seen in the energy drop between 10 and 20 ns, shown in red. A model of the lower energy structure is shown. Likewise, each newly created vacancy represents a higher energy state, which the graphene system can minimize through diffusion and incorporation with the existing defect cluster. Structural models of the defect cluster are shown for each incorporation event. What emerges is a chain of reconstructed divacancies aligned along the ZZ-direction, consistent with experimental observations. A video of the simulated structural evolution can be found in the SI.

We compare the accuracy of the structures obtained by the classical MD simulations with first-principles density functional theory (DFT) calculations using FHI-aims,[59] an all-electron code with localized numerical orbitals as the basis. We chose tight basis sets and Perdew-Wang local density approximation (PW-LDA)[60,61] as an exchange correlation function. The formation energies were calculated using both DFT and LAMMPS and are compared in Figure 4(d). Based on the structures identified via MD simulations, initial structures were created by manually modifying a hydrogenized 20 x 11 Å$^2$ graphene sheet (94 carbon and 26 hydrogen atoms, top and bottom edges are ZZ and the others are armchair). All structures were fully relaxed using the Broyden–Fletcher–Goldfarb–Shanno (BFGS)[62] algorithm until the maximum force component became less than $10^{-3}$ eV/Å. For LAMMPS, the structures were optimized until the maximum force became less than $10^{-8}$ eV/Å using the conjugate gradient (CG) scheme. Atomic models of each structure can be found in the SI. Figure 4(d) compares the formation energies from DFT and force-field potential using ARIEBO-M results. The energy difference between different structures is consistent, but with significant differences in the dangling bond states, which are over stabilized with the classical force field potential. The formation energies and optimum bond lengths of $C_2$ and $H_2$ obtained by the DFT and LAMMPS calculations are -9.5943 eV, 1.256 Å ($C_2$) and -6.6764 eV, 0.765 Å ($H_2$), and -6.2098 eV, 1.3255 Å ($C_2$) and -4.5063 eV, 0.746 Å ($H_2$), respectively.

Figure 4(e) shows the diffusion coefficient of a monovacancy as a function of temperature. To obtain this diffusion coefficient, we performed classical MD simulations. The interaction between carbon atoms in graphene was described by the AIREBO-M[57] potential implemented in LAMMPS.[58] We prepared a circular graphene sheet with a diameter of 200 Å containing 11345 carbon atoms in a 205x215x60 Å$^3$ periodic boundary cell. Then, we created a monovacancy by removing a center carbon atom of the graphene. During the simulation the outermost 5 Å edge atoms of graphene were fixed and we kept the system at a given temperature using NVT for a total simulation time of 10 ns.

Diffusion coefficients (D) were evaluated using the equation D = MSD/($2n\Delta t$), where MSD is the mean square distance, $n$ is the dimension of the sample, 2, and $\Delta t$ is the time interval. The MSD of a monovacancy at a given $\Delta t$ was averaged from 0.1 ns to 10 ns, the total simulation time, and D was obtained by fitting the MSD values as a function of $\Delta t$. We performed calculations for temperature ($T$) between 2000 K and 4000 K and obtained $D$ for the given $T$ (see the table in Figure 4(f)). Based on the Arrhenius equation, $D(T) = D_0 \exp(-\beta/(kT))$, we obtain $D_0$, the pre-exponential factor (nm$^2$/s), and the activation barrier for diffusion, $\beta$, where $k$ is the Boltzmann constant, also listed in the table.

In the calculation of MSDs, the position of a vacancy was calculated as the averaged coordinates of the carbon atoms with two or four neighbors with the distance between a C-C bond less than ~1.81 Å (0.68 Å covalent radii x 2 + 0.45 Å tolerance).

Interestingly, literature values for the predicted diffusion barrier of a monovacancy[63] differ significantly from that predicted by the AIREBO-M MD simulation suggesting it may be an overestimate. To examine what effect this has on the shape of the diffusivity curve with temperature we also plot the curve using the literature value of 0.94 eV. As expected, this increases the diffusivity at all temperatures. The important feature of these plots for the present purpose is not the precise values (and we make no claim to their absolute accuracy), rather it is the overall trend, which is governed by the Arrhenius equation. High temperature diffusivity is many orders of magnitude higher than it is at room temperature, which is in accord with the experimental observations. For example, at room temperature our MD simulations estimate it would take many years for a vacancy to diffuse a distance greater than the milling diameter. Using a diffusion barrier of 0.94 eV decreases this time to ten minutes. However, from the perspective of milling a hole in graphene these two scenarios are indistinguishable as the vacancy does not move significantly on the relevant time scale (10s of seconds). Likewise, the temperature at which a vacancy diffuses beyond the milling radius in one second is estimated at 271 ºC and 91 ºC for the MD simulation and the 0.94 eV barrier, respectively. This temperature represents roughly the point at which we expect diffusion to start playing a significant role in the milling process. Clearly, the MD simulation using AIREBO-M potential provides a rather conservative estimate for the vacancy diffusion process. Nevertheless, Airebo-M potential is able to reproduce the experimentally observed chains of atomic defects and to illustrate possible atomic structural transition pathways evaluated by DFT. In addition, when calculating the diffusion barrier of monovacancy using MD simulations, the Airebo-M potential overestimates the diffusion barrier compared to the DFT, but the calculated estimated temperature at which diffusion begins to play a significant role in the milling process is qualitatively consistent with our experimental observations.

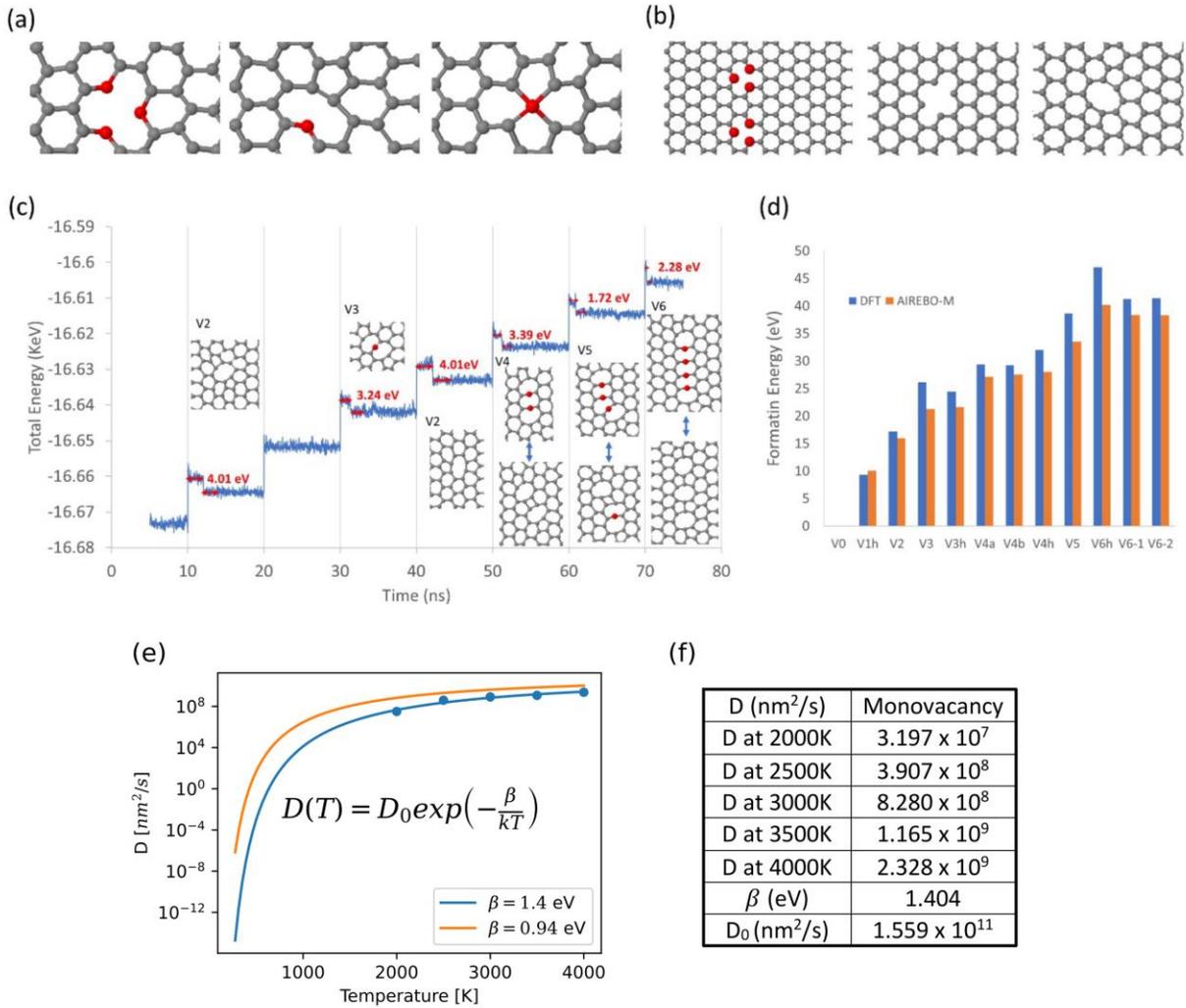

**Figure 4 MD simulations revealing the energetics of creating carbon vacancies and their kinetics to form topological defects in free-standing graphene.** (a) Diffusion process of a single vacancy. (b) Merging process of double vacancies. (c) Change of total energy during MD simulations. The structures where vacancies are merged are shown where red energy values are decreased energy values due to vacancy merging. Red arrows indicate where the energies were averaged and the averaged values were used to calculate corresponding decreased energy values (red). Blue arrows indicate reversible structure changes during MD. (d) Formation energies. (e) Diffusion rate as a function of temperature for a monovacancy through the graphene lattice. (f) Table listing the diffusion rates shown in (e).

With this understanding, the milling rate at high temperatures appears to be primarily governed by the vacancy diffusion rate relative to the vacancy generation rate. Vacancies will tend to diffuse away from the milling region. When two vacancies meet, they become fixed in the lattice. As more vacancies are added to the branching vacancy chain it expands until it finally reaches the milling region, at which point a stable hole can be formed. The details are clearly more nuanced, but this basic description captures the fundamental elements.

To examine how a system governed by these rules evolves, we created a simple computational model within which the rules were embedded. This model was run over longer time periods than what is accessible by MD simulations and we compare the outcome with experimentally observed statistical distributions. In this model, a regular (square) 2D array of potential vacancy sites represents the graphene at the atomic scale. A single vacancy is created at regular time intervals at the center of the array and allowed to diffuse in a random and unbiased manner, i.e., at each time step there is equal likelihood that the vacancy will move to any of its nearest neighbor locations or remain at its current location. If a defect diffuses to an edge of the array, it is eliminated from the simulation as it is considered to have diffused too far away to play any further role. As time progresses, more vacancies are generated. If any two of these vacancies become nearest neighbors through the random walk, they both are set to remain in their locations for the remainder of the simulation (i.e., have formed an immobile defect cluster). Likewise, if a single vacancy comes in contact with a fixed defect cluster or chain, it will also remain in place, increasing the size of the defect cluster. This simulation runs until the vacancy chain grows to the vacancy generation site at the center of the array (i.e., the beam location), at which point the simulation terminates with the assumption that the condition for hole milling is met. The simulation was run many times for each defect generation rate to determine the statistical distribution relating the time to form a hole (number of time steps) for that defect generation rate. In this model, the only adjustable parameter is the defect generation rate relative to the vacancy jump rate. Importantly, to relate this model to real world conditions, the ratio between the vacancy jump rate and the defect generation rate is used as a proxy for temperature, with the assumption being that the vacancy jump rate increases exponentially with temperature, but the vacancy generation rate remains relatively unchanged with temperature. Also note, no account is made for vacancies that spontaneously disappear, which occurs if they combine with a randomly diffusing carbon adatom. The likelihoods of vacancies meeting and not forming a stable defect complex or vacancies encountering a defect chain and not adding to it are not considered. Also, any diffusion biasing effects that may result from strain fields that are formed and change during defect chain creation are ignored. The Matlab code for this model is provided in a supplemental file.

Figure 5(a) shows a summary of the resulting simulation statistics generated. For each diffusion rate a histogram shows the number of steps required to build a defect chain extending to the defect generation position. As the diffusion rate is increased, the distribution skews to higher values. Each distribution was fit with a Birnbaum-Saunders model[71] to capture the overall shape where the optimum fit is shown by the black curves overlaid on the histograms. The Birnbaum-Saunders distribution was developed to model the fatigue-life of metals subjected to periodic stress[71] and has been applied generally for the modeling of accumulated damage.[72] As can be seen, this model nicely matches the shape of the distribution. After fitting each distribution, the curves were rescaled so that the maximum value was set to one and plotted together in Figure 5(b). Increasing the diffusion rate shifts more data points into the upper tail of the distribution until finally at a rate of 2000 the peak itself has also shifted upward. We note that most of the distribution shapes closely match at the beginning and the peak position remains fairly stable, thus making it difficult to observe these changes in the histograms. The hazard rate gives the instantaneous "failure rate," which is synonymous with a successfully milled hole. The hazard rate for each fit was calculated and is plotted in Figure 5(c). For the Birnbaum-Saunders distribution these hazard rates asymptotically converge to a single value indicated by the dotted lines in Figure 5(c) and are replotted for clarity in Figure 5(d) as a function of the diffusion rate. Here, we can clearly see that the increased diffusion rate has a pronounced effect on the long term behavior of the system.

These simulated results were compared with the experimental observations. In Figure 5(e) histograms of the dose required to mill a hole at each temperature are plotted. Because the beam current was kept approximately constant, the dose also corresponds to elapsed time. The increase in temperature drastically increases the vacancy diffusion rate, as shown in Figure 4(e), so the temperature also corresponds to the diffusion rate (inverse of time step) in the simulation. Birnbaum-Saunders distributions were fit to this data and a normalized comparison of each fit is shown in Figure 5(f). For the 1000 ºC case, only 25 holes were attempted, 11 of which were successful (shown) and 14 of which were not successfully milled after an accumulated dose of $180 \times 10^9$ e/nm$^2$. These failed attempts were excluded from the fit. Thus, the 1000 ºC case does not contain enough successful milling attempts to be meaningfully captured by the fit and the majority of the data suggested even higher doses would be required, which are not included in the fit. This data is included for completeness. The corresponding hazard functions are shown in Figure 5(g) and the asymptotic limit values are plotted in Figure 5(h).

While this simple model only qualitatively captures the effect of varying diffusion rates and the idea that vacancies diffuse while divacancies remain fixed, this is sufficient to qualitatively reproduce the experimentally acquired milling data. The overall shape of the statistical distribution is similar, the change in the shape of the distribution with increasing diffusion rate is similar, and the trend in the hazard functions is similar. This suggests that modeling of such processes may be possible without recourse to the full atomistic physical description of the system and future efforts may provide fruitful inquiry in this direction. Moreover, the parsimonious description of the physical mechanisms governing the milling process outlined above are sufficient to qualitatively reproduce the behavior of the system. This suggests that the most important factors have been identified.

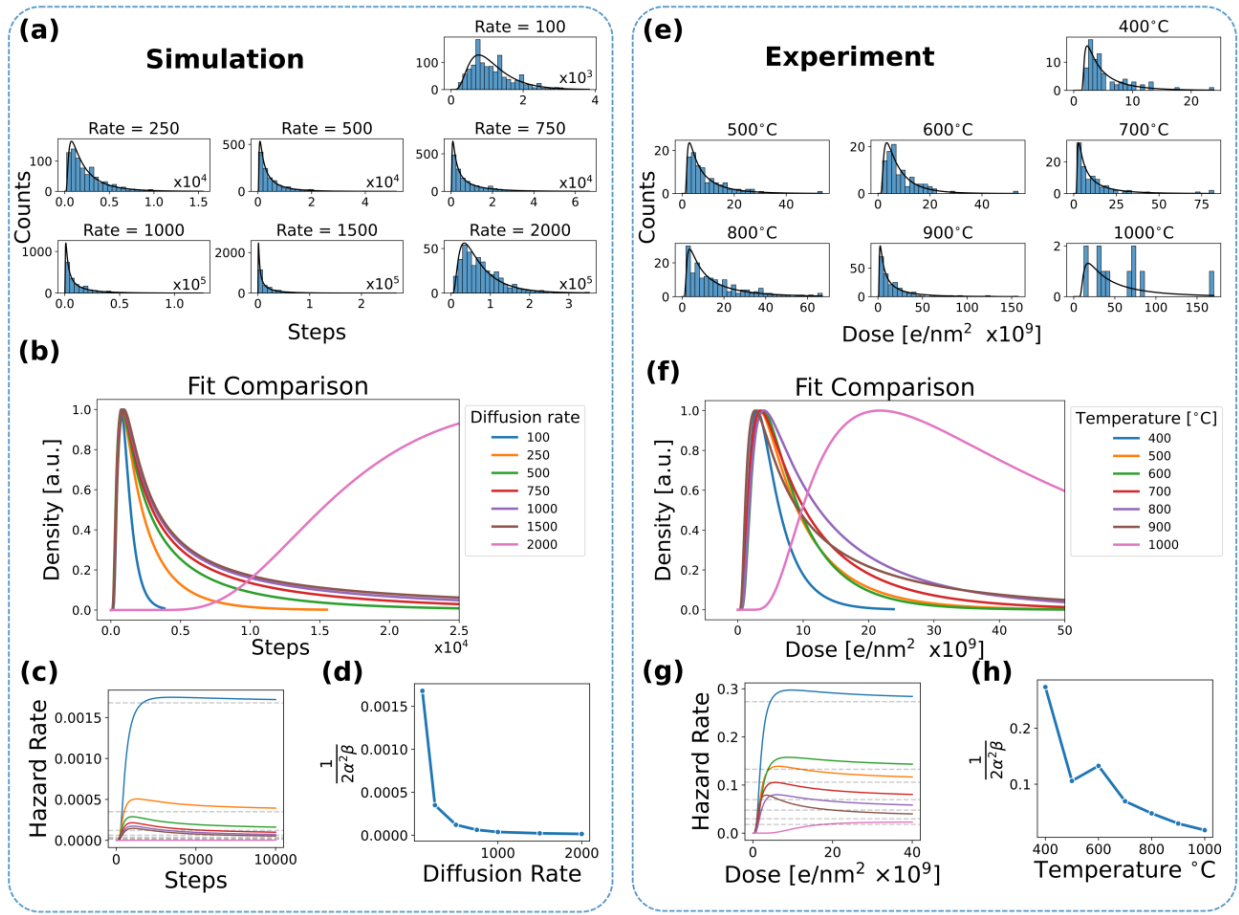

**Figure 5 Summary of simplified hole milling model and qualitative comparison with experiment.** (a) Histograms of the number of steps required to form a hole at various defect generation rates. Birnbaum-Saunders distributions were fit to each dataset (black curves), rescaled so that the maximum equaled one, and plotted for comparison in (b). The hazard function for each curve is shown in (c) and the asymptotic limit values are shown in (d). (e-h) Similar analysis for the experimental data acquired at each temperature.

## 4 Conclusion

We explored the dynamic processes of defect diffusion and healing in graphene over a range of elevated temperatures during e-beam irradiation above the knock-on threshold. An apparent resistance to e-beam damage during imaging was observed as the graphene temperature was increased. This counterintuitive observation can be largely explained by the increased diffusion rate of defects at elevated temperatures. Although vacancies continue to be generated at high rates by the e-beam, they quickly leave the irradiated region and temporarily avoid local agglomeration into vacancy clusters (i.e., extended restructured defect clusters residing close to the point of generation). As the temperature is increased and the diffusion rate increases, vacancy/defect clusters have a higher probability of nucleating farther from the e-beam. This phenomenon presents itself as an apparent increased resistance to local radiation damage.

This work helps define appropriate ambient conditions under which e-beam based modification of graphene can be performed successfully. These insights are also valuable for extension to other 2D materials by narrowing the relevant physical phenomena governing structural evolution and lend insights into bulk layered materials. For example, the behavior of defects in graphite is especially important with regard to degradation rates and modes in graphite reactor components. Thus, monolayer or few-layer graphene represents a powerful experimental platform and approach to study defect generation and the dynamics of defect accumulation with real-time *in situ* atomic resolution imaging.

## Acknowledgements

This work was supported by the U.S. Department of Energy, Office of Science, Basic Energy Sciences, Materials Sciences and Engineering Division (O.D., S.Y., M.Y., A.R.L, S.J.) and was performed at the Oak Ridge National Laboratory's Center for Nanophase Materials Sciences

**Supplemental Information**

**Figure S1** shows overview images of the suspended graphene membrane before, (a), and after, (b), heating. The darker areas in (b) are regions of clean graphene where the milling experiments were performed.

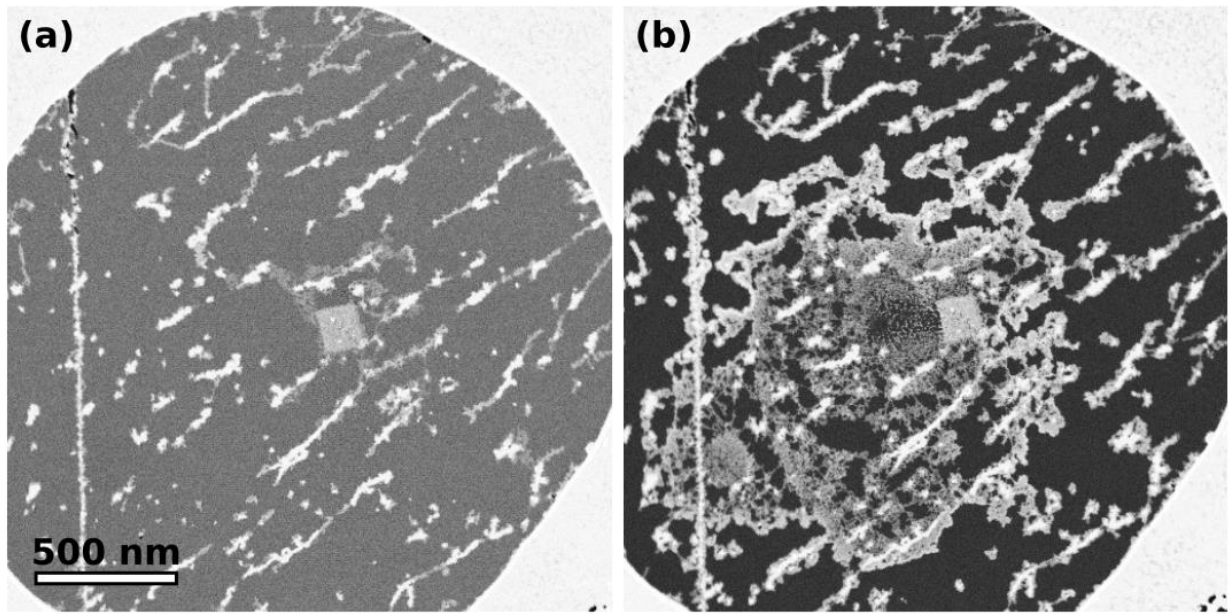

**Figure S1** Overview images of the suspended graphene membrane at room temperature before heating, (a), and after heating (b).

**Figure S2** shows several frames from a 45-frame image sequence acquired at room temperature with a 100 kV e-beam. The frames were acquired sequentially resulting in images of graphene through a progressively increasing total electron dose.

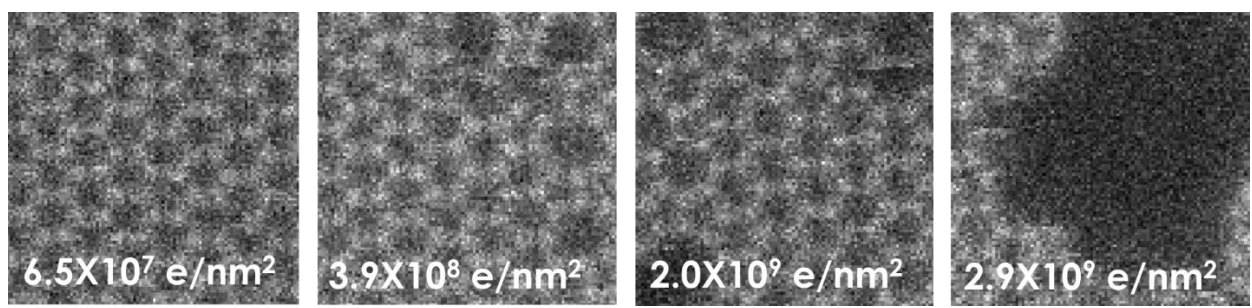

**Figure S2** Image sequence of the state of a graphene lattice after exposure to increasing levels of 100 kV electron irradiation at room temperature. Total accumulated dose is listed at the bottom of each image. The field of view is 1.6 nm.

In Figure 4(d) of the main text we showed a comparison of DFT and AIREBO-M optimized structures, reproduced in to top panel of **Figure S3**. In the lower panel we show the associated structures for each calculation. The labels give additional information: the number indicates how many carbon atoms have been removed, the "h" indicates where hydrogen has been introduced to passivate dangling bonds, and the "a" and "b" indicate different possible structures which were both observed in the simulation.

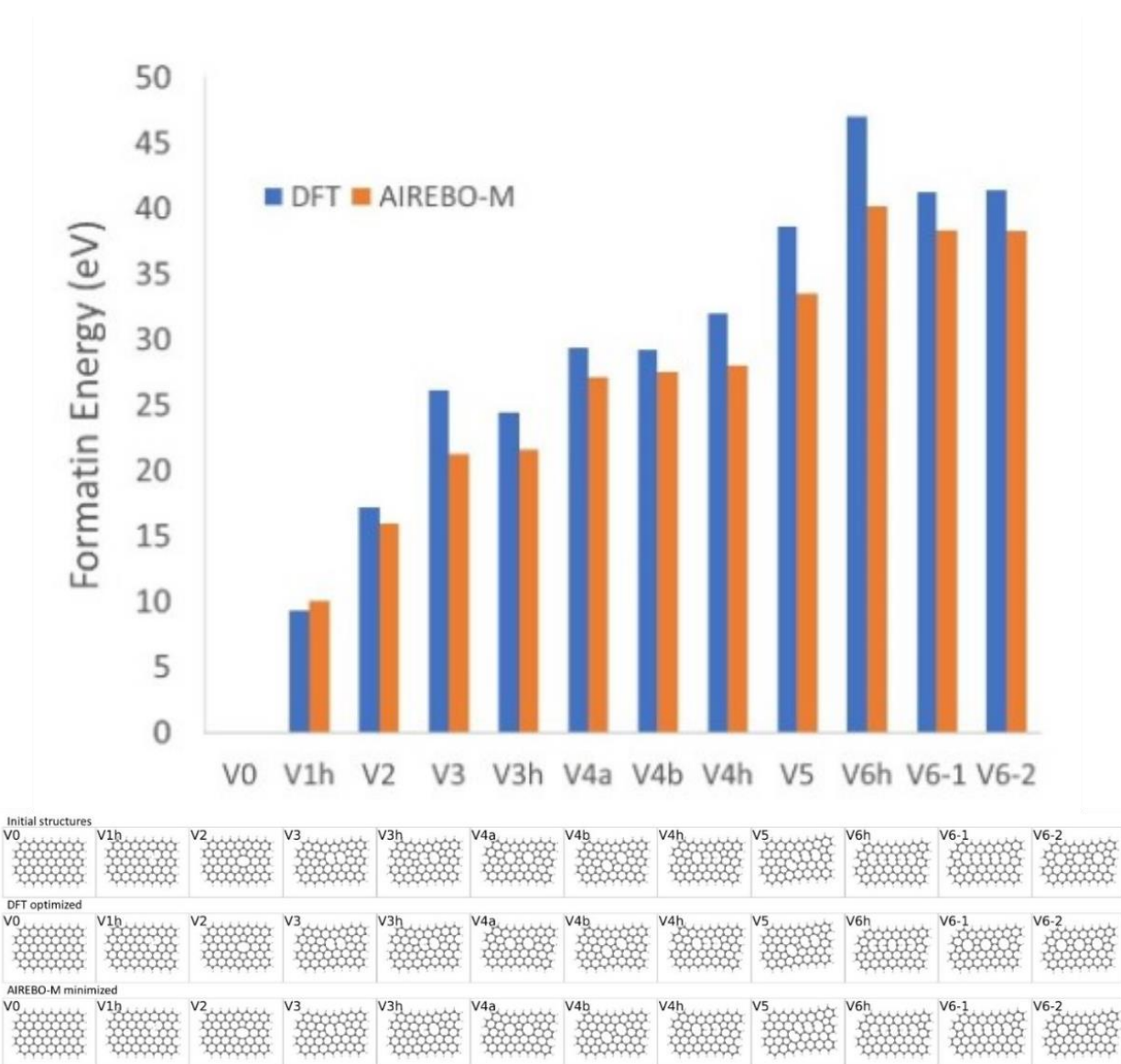

**Figure S3** Comparison of the optimized structures calculated in Figure 4(d) of the main text. Figure 4(d) is reproduced here in the top panel for reference. The associated initial and optimized structures are shown below.